# White Paper: Recommendations for immersive accessibility services

## Editors

- *Peter tho Pesch (IRT, Germany), peter.thopesch@irt.de*
- *Romain Bouqueau (Motion Spell, France), romain.bouqueau@motionspell.com*
- *Mario Montagud (i2CAT, Spain), mario.montagud@i2cat.net*

## Contributors

This White Paper has been prepared by the ImAc project (https://www.imac-project.eu/) consortium, which includes the following nine partners (see Figure 1):

- I2CAT Foundation (Spain) - Project Coordinator.
- CCMA (Spain)
- Anglatècnic (Spain)
- UAB (Spain)
- IRT (Germany)
- RBB (Germany)
- Motion Spell (France)
- RNIB (UK)
- University of Salford (UK)

## Abstract

This paper provides recommendations on how to integrate accessibility solutions, like subtitling, audio description and sign language, with immersive media services, with a focus on 360º video and spatial audio. It provides an in-depth analysis of the features provided by state-of-the-art standard solutions to achieve this goal, and elaborates on the finding and proposed solutions from the EU H2020 ImAc project to address existing gaps. The proposed solutions are described qualitatively and technically, including example implementations. The document is intended to serve as a valuable information source for early adopters who plan to provide accessibility services to their portfolio with standard-compliant solutions.

Figure 1. ImAc project consortium (https://www.imac-project.eu/))

# Table of contents

# 1. Introduction & Motivation

ImAc (Immersive Accessibility) is an EU H2020 project (Oct 2017- Mar 2020) that has explored how to efficiently integrate accessibility solutions, like the personalized presentation of access services, with immersive media, focusing on 360º video.

Augmenting 360º video services with accessibility features requires the addition of potential extensions to the delivered streams and media formats, which are not yet covered by current standards. While implementing proprietary solutions to be adopted in closed environments becomes an alternative, this was not the goal of ImAc, as one of the key objectives of the project has been to work towards providing interoperable and standard-compliant solutions.

This document provides an overview of findings, lessons learned and contributions of the project in terms of proposed extensions to current standard delivery technologies and media formats to achieve the targeted goals. Therefore, the document is intended to serve as a valuable information source for early adopters who plan to provide accessibility services to their portfolio with standard-compliant solutions.

This document provides key insights from a thorough analysis of existing standards, mainly MPEG DASH for media delivery and TTML for subtitles and related metadata, to assess how additional metadata can be incorporated to meet the gathered accessibility requirements (from conducted user-centric activities and user testing). It was found that the requirements and necessary extensions for each access service were quite diverse, ranging from proposing extensions to current standard signaling features (e.g. for subtitles) to having to use new ad-hoc metadata levels (e.g. for sign language). The key findings for each access service are summarized next:

- **Subtitles**: Using the Internet Media Subtitles and Captions IMSC (https://www.w3.org/TR/ttml-imsc1.0.1/) format for subtitles, additional data can be carried within the subtitle format. The adaptation of the subtitle service to a 360° environment requires reconsidering the positioning strategies of subtitles (refer to section 2.4.2), as well as the reference to the spatial position of the related speaker (refer to section 2.4.3). An exception applies though: signaling solutions to indicate the availability of "easy-to-read" subtitle variants need to be provided, and this should happen at the MPEG DASH layer, but it is not currently specified in the standard.
- **Audio Description and spoken subtitles**: The ImAc specific extensions regarding audio description and spoken subtitles are processed and pre-rendered during the production process at the server side. In such services, relevant parameters about the various available audio streams (e.g. relative gain, spatial audio mode, scripting style...) need to be signalled in order to enable the player application to identify them, and thus enable the user selection. In this context, dedicated attributes within the DASH format have been proposed to identify the different ImAc variations for these services (refer to section 3.3.2 and 4.1). An additional gap is the absence of any standardized solution for identifying an audio stream as Ambisonics (a coding for spatial audio). This is a requirement that we believe others will have as well, and thus we have filed a corresponding proposal to MPEG (refer to section 3.3.1). Likewise, the simultaneous presentation of more than one audio stream at a time may also require modifications to the adopted media player.
- **Sign Language**: Unlike the available Sign Language services which are provided as burned-in videos, ImAc has provided a more interactive and customisable solutions with the use of an independent stream for the sign language video, with some enriched metadata (e.g. position of the target speaker, descriptive textual information, signaling of on/off periods based on the interpreter's activity…). Enabling all these features has been more challenging and required extra extensions, which consisted of the definition of a sidecar (accompanying) metadata TTML file carrying all relevant information, with the proper timeline. The proposed solution for embedding the metadata file in the MPEG DASH manifest is described in section 5.3.1. The separate data fields we introduced are described in sections 5.3.2 – 5.3.5.

# 2 Overview of Requirements and Gap Analysis

| Requirement / Challenge | Description and current status |
|---|---|
| Easy-to-read-subtitles | *ImAc feature* - Easy-to-read subtitles are an alternative subtitle service / track that may be present in addition to the standard subtitle track (in the same language). Besides, it is also possible that the (only) subtitle track for a language may meet the conditions of "easy-to-read", and thus would need to be labelled accordingly.<br>*Proposed solution* - The signaling must be available to the player, ideally without the need to decode each available subtitle track. Typically, a player would show the available subtitle services on a UI to the user, such that the preferred option can be selected.<br>This information cannot be provided by the current standards relevant for the ImAc services. To enable the above mentioned player functionality, the signaling must be provided on the MPEG DASH level. |
| Speaker position (subtitle) | ImAc feature – ImAc suggests adding visual indicators that show the position of the current speaker. Such an indicator may be an arrow that is shown next to the subtitle. An alternative subtitle presentation mode is to render the subtitle next to the speaker. For both options, the speaker position must be known.<br>*Proposed solution* -In MPEG OMAF, a position within a 360° scene can be related to a subtitle. However, this position describes the position of the subtitle itself and not the position of the related speaker. There is currently no mechanism to express the related speaker position (or direction) in a standardized way.<br>The IMSC subtitle format allows the usage of custom attributes and may be a candidate to store this information. |
| Audio description gain | ImAc feature – The user may choose between three different audio mixes that differ regarding the Audio Description (AD) gain in relation to the gain of the main audio.<br>*Proposed solution* -There is currently no mechanism in the standards used for ImAc content distribution to signal this information. By design, each mix results in a list of adaptation sets in the MPEG DASH manifest (at least one, more if several quality levels are provided). Thus, the adaptation set element is a good candidate position to signal this information. |
| Audio description mode | ImAc feature – The user may choose between different treatments of the AD audio signal (AD modes).<br>As for the AD gain, there is currently no mechanism in the standards used for ImAc content distribution to signal this information. By design, a different audio treatment results in an additional mix, which results in a list of adaptation sets in the MPEG DASH manifest (at least one, more if several quality levels are provided). Thus, the adaptation set element is a candidate position to signal this information. |
| Extended Audio Description | ImAc feature - One of the gathered requirements in ImAc has been to enable Extended AD features, in which beep notifications are provided to the users when additional AD information is available. If the user wishes to receive this info, then the main video and AD tracks are paused, and an associated Extended AD track is presented. Once reached the end of the clip, the playout resumes.<br><br>There is absolutely no support for this feature in state-of-the-art solutions and standards, so ImAc has provided an innovative solution based on extending the attributes of the Adaptation Sets to be able to link a set of Extended AD tracks to main AD track(s), providing the relevant metadata. |
| Audio channel description | ImAc feature - ImAc makes use of spatial audio for both the main audio mix as well as the audio description part. The format used in the project is ambisonic.<br><br>*Proposed solution* - In MPEG DASH the signaling of the audio channel allocation is made through the AudioChannelConfiguration attribute and the @schemeIdUri attribute "urn:mpeg:dash:23003:3:audio_channel_configuration:2011". For instance, this scheme describes where the channel of the center speaker of a 5.1 system is located in the stream. For Dolby encoded streams, there's another scheme defined using the schemeIdUri "tag:dolby.com,2014:dash:audio_channel_configuration:2011"<br><br>However, it is not possible to indicate any format that is NOT loudspeaker-based. That means formats without a 1:1 mapping between an audio channel and a loudspeaker. This is for example the case for Ambisonics which is transported in four audio channels but need to be decoded properly before mixed for instance to a headphone signal. |
| Sign language track (closed) | The Sign Language (SL) video is provided as an independent track, and its signaling will be supported by an upcoming version of MPEG-DASH (currently under draft). But by itself, the MPEG DASH specification is not enough for a harmonised implementation. Additional usage guidelines like the DVB DASH profile are required.<br><br>Likewise, rendering information to combine a closed SL video with the main video in the player cannot be signaled in MPEG DASH (e.g. positioning information like location or size – needed to render the video from an additional adaptation set onto the main video).<br><br>In addition, most media players, like dash.js, only support the playout of one video track and one audio track. ImAc has provided a solution to enable the presentation of >1 video stream at a time. |
| Sign language track status indication | ImAc feature – Only show sign interpreter video when the sign interpreter is translating.<br><br>A status indication for the video indicating whether the sign interpreter is currently translating is not supported.<br><br>In MPEG DASH, multi-period Media Presentation Descriptions (MPDs) and separate tracks for the signer video segments are possible. However, having a multi-period signer video would also involve having a multi-period video for the main 360º video. In addition, the |
| | current DASH client implementations showed limited compatibility due to vagueness in the standards. Finally, manifests would become bigger when sign language is often switched on and off, which was not ideal for the broadcasters.<br><br>*Proposed solution* -The use of an accompanying TTML file to provide the required (textual and temporal) metadata is proposed in ImAc. |
| Speaker position (signer) | ImAc feature – Speaker identification for signer videos are supported by adding a graphical element (e.g. arrow) to the scene that guides toward the position of the (current) speaker.<br><br>Neither MPEG4 nor MPEG DASH support a mechanism to add such metadata to the signer video track.<br><br>The use of an accompanying TTML file to provide such metadata is proposed in ImAc. |
| Speaker identification (signer) | ImAc feature – Speaker identification for signer videos are supported by adding a speaker's name or an emoji under the signer video or a coloured frame around it.<br><br>Neither MPEG4 nor MPEG DASH support a mechanism to add such metadata to the signer video track.<br><br>The use of an accompanying TTML file to provide such metadata is proposed in ImAc. |
| Audio Subtitles (AST) presence (delivered as audio) | Spoken subtitles can be realised either by a text-to-speech process in the client device (using a subtitle track), or alternatively, by delivering a pre-rendered audio track that contains spoken subtitles. When using the latter approach, the audio track must be signaled accordingly, such that the receiver knows the content of this stream. The content may be either AST only, or AST pre-mixed with main audio or even AD and main audio.<br><br>In ImAc, AST is delivered as a separate audio track and mixed with the main audio at the player side, if selected.<br><br>MPEG DASH supports the signaling of an additional complementary audio track (AST falls into this category) but doesn't support any means to identify the stream as AST. Likewise, most media players, like dash.js, only support the playout of one video track and one audio track. ImAc has provided a solution to enable the presentation of >1 audio stream at a time. |
| AST mode | ImAc feature – The user may choose between different treatments of the AST audio signal (AST modes).<br><br>There is currently no mechanism in the standards used for ImAc content distribution to signal this information. By design, a different audio treatment results in an additional mix, which results in a list of adaptation sets in the MPEG DASH manifest (at least one, more if several quality levels are provided). Thus, the Adaptation Set element is a candidate position to signal this information. |

# 3 Technical recommendations

## 3.1 Subtitles

### 3.1.1 Motivation

At the beginning of the ImAc project, we gathered a list of user requirements for presenting subtitles in 360° videos derived from conducted user-centric activities. Based on these requirements, different presentation modes were implemented and evaluated via various user testing phases (described in detail in D5.2 [3]).

We found that current standards and specifications do not provide the required possibilities to realize all features of the developed presentation options. In this document, we describe possible solutions in detail, based on existing standards.

In particular, we have identified the following requirements that arise from the ImAc use cases[1]:

1. Means to signal a subtitle service as "easy to read"
2. Means to position subtitle regions on a 2D plane in the field of view of the user.
3. Means to locate the audio source of the timed text horizontally (e.g. with a longitude coordinate).

Sections 3.2 and 3.3 describe relevant data for subtitles in the state-of-the-art MPEG DASH and IMSC standard specifications.

Proposals for possible solutions to meet these requirements, based on DASH and IMSC, are described in Section 3.4.

In addition, the project has considered an additional presentation mode for subtitles, which consists of displaying them next to the speaker, fixed to the 360° scene, and use always-visible indicators to guide the users towards the active speakers. Conducted tests did not confirm a preference towards this mode, but opened the door to further investigations about their potential applicability and benefits.

### 3.1.2 State of the art – TTML subtitles in DASH

IMSC subtitles can be signaled and described in MPEG DASH [4] using the DVB-DASH profile [5]. This is summarized in the following subsections.

#### 3.1.2.1 Role element

**Location:**
AdaptationSet/ContentComponent elements - as part of an MPD

**Defined in:**

- ISO/IEC 23009-1:2014 [4], § 5.8.4.2, § 5.8.5.5
- ETSI TS 103 285 V1.1.1 [5], § 7.1.2

[1] The use cases that were tested with end users during various test phases are summarized in D5.2, second iteration [3]. Features for subtitles are listed in chapter 3.1.1.1, features for easy-to-read subtitles are listed in chapter 3.1.1.2 of this document.

**Description:**
The Role element is an element in DASH to specify an arbitrary role for the affected component that is based on a specific scheme. The element can be used multiple times on a component.

**Role element – @schemeIdUri attribute**
This attribute specifies the scheme used for the role value. The DASH specification itself defines a simple role scheme (URI: urn:mpeg:dash:role:2011) which shall be used here.

**Role element – @value attribute**
This attribute specifies the actual role value. Regarding subtitles, the applicable values depend on the actual purpose of the subtitles. The following values from the mentioned scheme are currently used:

- main
- alternate
- commentary

#### 3.1.2.2 Accessibility element

**Location:**
AdaptationSet/ContentComponent elements - as part of an MPD

**Defined in:**
- ISO/IEC 23009-1:2014 [4], § 5.8.4.3
- ETSI TS 103 285 V1.1.1 [5], § 7.1.2
- ETSI TS 102 822-3-1 V1.9.2 [6], § A.15

**Description:**
The Accessibility element is a generic element in DASH to specify an arbitrary accessibility property that is based on a specific scheme. The element can be used multiple times on a component. Regarding subtitles this element is only used for subtitles for the hard of hearing.

**Accessibility element – @schemeIdUri attribute**
This attribute specifies the scheme used for the accessibility value. DASH itself does not define a scheme here. For subtitles for the hard of hearing, the *Audio Purpose* scheme from the *TV Anytime* project is used (URI: urn:tva:metadata:cs:AudioPurposeCS:2007).

**Note:** It is indeed correct that a scheme actually designated for audio is used for subtitles here.

**Accessibility element – @value attribute**
This attribute specifies the actual accessibility value. To indicate subtitles for the hard of hearing, the value 2 from the mentioned scheme is used here.

#### 3.1.2.3 AdaptationSet element

**@mimeType attribute**

**Location:**
AdaptationSet/Representation/SubRepresentation elements - as part of an MPD

**Defined in:**
- ISO/IEC 23009-1:2014 [4], § 5.3.7.2
- ETSI TS 103 285 V1.1.1 [5], § 7.1.1

**Description:**
The mandatory mimeType attribute is used to describe the MIME type of the actual media segments.

For TTML subtitles, the value "application/mp4" (TTML encapsulated in ISOBMFF) is used.

**@codecs attribute**

**Location:**
AdaptationSet/Representation/SubRepresentation elements - as part of an MPD

**Defined in:**
- ISO/IEC 23009-1:2014 [4], § 5.3.7.2
- ETSI TS 103 285 V1.1.1 [5], § 7.1.1

**Description:**
The optional codecs attribute specifies the used codec. It shall also include profile and level information, if applicable. For TTML subtitles, the value shall begin with "stpp" indicating XML subtitles. The value should indicate the used profile according to the W3C TTML profile registry. If the value "stpp" is used, the value "stpp.ttml.etd1" shall be assumed, referring to TTML content suitable for an EBU-TT-D renderer.

**@lang attribute**

**Location:**
AdaptationSet/ContentComponent/ProgramInformation elements - as part of an MPD

**Defined in:**
- ISO/IEC 23009-1:2014 [4], § 5.3.3.2
- ETSI TS 103 285 V1.1.1 [5], § 7.1.2

**Description:**
The lang attribute is used to describe the actual language of e.g. a subtitle adaptation set. Its value is specified according to RFC 5646 [18].

### 3.1.3 State of the art – Positioning of subtitles in IMSC

The following description of positioning subtitles within the IMSC root container assumes the writing direction left to right and top to bottom. Note that the writing direction of a subtitle language may affect the positioning of subtitles within a region.

#### 3.1.3.1 Root container region

**Location**
n.a.

**Defined in**

- W3C IMSC 1.0.1 [7], §6.7.1
- W3C TTML1 [8], §2.2

**Description**
The Root Container Region of a Document Instance is mapped to each image frame of the Related 2D Video Object. The presence of the attribute aspectRatio may result in a root container region that does not match the image frame in its entirety.

#### 3.1.3.2 Region element

**Location**
/tt/head/layout

**Defined in**

- W3C IMSC 1.0.1 [7], §6.8.2
- W3C TTML1 [8], §9.1.2

**Description**
Defines a space or area for the display of subtitle content. The region shall not extend the root container region.

**@origin attribute**
The origin attribute sets the offset of the region of the top left corner of the root container region.

**@extent attribute**
The extent attribute sets the width and height of the region.

#### 3.1.3.3 Style element

**Location**
/tt/head/styling

**Defined in**

- W3C IMSC 1.0.1 [7], §7.4
- W3C TTML1 [8], §8.1.2

**Description**
The styling element contains a set of style information through style attributes. Among others, the (horizontal) alignment of subtitle blocks is defined via the tts:textAlign attribute.

**@textAlign attribute**
It defines the alignment of a subtitle block. Allowed values are “left”, “center”, “right”, “start”, “end”

### 3.1.4 Proposed solutions for ImAc services

As described in Section 3.1, we identified the following features that are not sufficiently supported by relevant standards:

1. Means to signal a subtitle service type as “easy-to-read”
2. Means to position subtitle regions on a 2D plane in the user’s field of view.
3. Means to locate the audio source of the timed text horizontally (e.g. with a longitude coordinate).

The following subsections describe possible solutions to realize these features.

#### 3.1.4.1 Easy to read subtitles

Easy-to-read subtitles are an alternative subtitle service / track that may be present in addition to the standard subtitle track (in the same language). Alternatively, the (only) subtitle track for a language may meet the conditions of “easy-to-read” and may be labelled accordingly.

Regarding the meaning of easy-to-read, the ImAc project refers to the description in ITU-R BT.2207-3 [9], Chapter 6. In short, easy-to-read subtitles is a broadcast service tailored to the language comprehensive level of non-native speakers, children, people with developmental impairments and people with cognitive losses, due e.g. to aging.

The signalling for easy-to-read must be available to the player, ideally without the need to decode each available subtitle track. Typically, a player would show the available subtitle services on a User Interface (UI) to the users, such that they can select the preferred option.

**Proposed solution**

Following this requirement, we propose to signal easy-to-read at the DASH layer (or equivalent when using other streaming formats, like HLS or CMAF). This value shall identify the labelled media as intended for people who need a simple language to be able to follow the content. The labelled media should follow the recommendations for easy-to-read language described in ITU-R BT.2207-3 [9], Chapter 6.

For the usage of MPEG DASH, we propose to add an easy-to-read identifier as a new value to the descriptor “urn:tva:metadata:cs:AudioPurposeCS:2007” for the Accessibility element of the MPD. This descriptor uses integer values that are mapped to target audience groups. For example, the value “2” is used to indicate the target group “for the hard of hearing”. The format of such a new value must still be discussed.

An alternative solution would be to signal easy-to-read in the subtitle format, in our case IMSC. But a player would need to decode all subtitle files in order to determine if a track contains easy-to-read subtitles. Although this is a simple readout, it would be an unnecessary task in the player.

**Additional proposal for the TTML file format**

To prepare the subtitle file for the correct signalling in distribution, the subtitle file should be labelled accordingly during the authoring stage. One possibility would be to add this as an additional specific content type for “Easy to Read” in the EBU-TT Classification Scheme [10] used in the ebuttm:documentContentType element, which is used in EBU-TT to set the content type for certain subtitle content.

A sample implementation of this feature in ImAc is provided in Annex I.

#### 3.1.4.2 Positioning of a subtitle region in 360° scenes

The positioning means of IMSC do not currently include the use case of 360° video environments, and consequently do not allow defining the position of the subtitle region in a 360° video or even a 3D scene.

Two general approaches for subtitle positioning were tested in the ImAc project:

1. Rendering subtitles always visible fixed to the screen / viewport (i.e. field of view)
2. Rendering subtitles fixed to the video (e.g. attached to the speaker)

Our proposal refers to the first approach (subtitles always visible fixed to screen). As described in Section 3.1.

The IMSC specification (W3C IMSC 1.0.1 [7]) requires the allocation of a “root container region” that establishes a 2D coordinate system where regions can be rendered into. IMSC further specifies that the root container region spans over the entire image frame (special rules apply if the attribute ittp:aspectRatio is set though).

Following the TTML specification, the properties of the root container region may be established by the document processing context. But such a processing context has yet to be defined for a 360° scene.

We found that neither TTML itself, nor any of the TTML-derived formats (like IMSC), provides a 2D coordinate system within the 360° scene that may be referred to as the root container region.

**Proposed solution**

We will describe a possible mapping of a 2D coordinate system into a 360° scene. This is essential for interlinking a 2D subtitle representation with a 3D space.

***Coordination system mapping*** – This section describes the mapping of a 2D coordinate system into a 360° scene. In this context, we understand the 360° scene as a 3D space (spanned by X-, Y-, and Z-axis) with the following constraints:

- The 3D space's purpose is to render an image for the presentation on a screen. The screen may be a tablet or PC screen, or the (stereoscopic) screen of an HMD.
- The image is retrieved by a rendering process that calculates the picture that a virtual camera would capture. The virtual camera is located at the origin of the coordinate system that spans the 3D space. The camera may be stereoscopic. The rotation of the virtual camera and its optical parameters define the part of the scene that is visual on the retrieved image.
- The 3D space further contains a non-stereoscopic video mapped onto a geometrical shape, which center of mass lies at the origin of the coordinate system that spans the 3D space. Typically, the geometrical shape is a sphere or a cube, depending on the representation format / encoding of the 360° video images.

In the 3D scene with the properties described above, a 2D coordinate system is spanned that represents the root container region of a TTML document. The plane spanned by the 2D coordinate system with the axis X and Y shall have the following properties:

- The plane shall lie between the origin of the 3D coordinate system and the video's geometrical shape, but may intersect with this shape.
- The plane shall lie orthogonal to the vector that goes through the origin of the 3D coordinate system and the center of the current viewport. The point where this vector intersects the 2D plane is also the origin of the 2D coordinate system.
- One of the axes (typically the X-axis) of the 2D plane shall be parallel to the horizontal plane spanned by the 3D coordinate system.

As a result, the subtitle plane will stay fixed to the screen and will always be visible to the user.

**Note**: In the stereoscopic rendering, the depth of the subtitle plane within the 360° scene affects the perceived image on the HMD. The distance of the subtitle plane from the virtual camera is left to the player implementation. However, it is suggested to put the subtitle plane as close as possible to the geometrical shape the video is projected on.

***Mapping of an equirectangular video texture*** – An equirectangular video texture is mapped onto a video object in the shape of a sphere in the 3D coordinate system. The texture shall be mapped to a sphere such that the positive X-axis of the 3D coordinate system intersects the center of the video texture.

#### 3.1.4.3 Location of the audio source

**Short summary of ImAc's subtitle presentation options**

Information on the (horizontal) location of the audio source is required to realize the enhanced subtitle presentation options developed in the project. The enhancement comprises two different ways of indicating the speakers position in the 360° scene for the active subtitle. The indication is done in form of a visual cue for which two options have been developed:

a) Arrow indicator – An arrow is placed left or right next to the subtitle, pointing into the direction of the related speaker, if the speaker is currently not visible within the picture. The arrow disappears as soon as the speaker becomes visible (i.e. moves into the picture) and appears as soon as the speaker disappears.
b) Radar indicator – a radar system is rendered next to the subtitle (e.g. on the right side) showing the current viewing direction and viewport. The radar is always active. For each active subtitle, a colored mark matching the subtitle color is rendered into the radar system indicating the horizontal position of the related speaker within the 360° scene.

**Proposed solution**

The subtitle presentation options proposed in the ImAc project require the information about the (horizontal) position of the related speaker in order to render a visual cue in form of an arrow or a radar system, as described above.

The information is required for each timed text segment, and thus can be transported on the subtitle format level. This is the simplest way to identify and describe separate timed text (subtitle) segments.

We propose to add attributes to the IMSC format that describe the horizontal and vertical direction of an audio source. The attributes should be considered to be significant only when applied to a tt:p or a tt:span element. The information may be used by the player application to provide graphical or other indicators for the position of the speaker in the scene. The speaker direction is represented by two angles, based on the coordinate system described in section 1.5.2, as follows.

The direction is given by two angles: azimuth and elevation. For each angle an attribute is added under the 'imac' namespace:

- Namespace used: http://www.imac-project.eu
- Abbreviation used: imac

The direction of the positive X-axis of the 3D coordinate system represents the angles azimuth=0° and elevation=0°. Increasing elevation angles refer to an upwards direction (i.e. positive values refer to a direction pointing above the horizontal X- Y-plane). Azimuth angles increase clockwise when looking from the center of the 3D space.

The attribute imac:audioSourceAzimut describes the azimuth (horizontal direction) of an audio source. The attribute should be considered to be significant only when it is applied to a tt:p or a tt:span element. The range of the azimuth is [-180, 180].

The attribute imac:audioSourceElevation describes the elevation (vertical portion) of the direction of an audio source. The attribute should be considered to be significant only when it is applied to a tt:p or a tt:span element. The range of the elevation is [-90, 90].

Refer to Annex I for a sample from the ImAc implementation.

## 3.2 Audio Description

### 3.2.1 Motivation

In section 2, we provided a gap analysis regarding the signalling of ImAc services and an overview of the required ImAc extensions. In this section, we describe possible solutions in detail, based on existing standards. Section 4.2 describes relevant data for audio and audio description, focusing on the state-of-the-art MPEG DASH specification.

The requirements for audio description (AD) that arise from the ImAc services are:

1. Means to signalize an Ambisonics audio stream in MPEG DASH.
2. Means to signalize custom properties of audio streams in MPEG DASH.

Proposals for possible solutions to meet these requirements, are described in Section 4.3.

### 3.2.2 State of the art

Audio Description can be signalled and described in MPEG DASH using the DVB-DASH profile, as summarized in the following sub-chapters.

#### 3.2.2.1 Role element

**Location:**
AdaptationSet/ContentComponent elements - as part of an MPD

**Defined in:**

- ISO/IEC 23009-1:2014 [4], § 5.8.4.2, § 5.8.5.5
- ETSI TS 103 285 V1.1.1 [5], § 6.1.2

**Description:**
The Role element is an element in DASH to specify an arbitrary role for the affected component that is based on a specific scheme. The element can be used multiple times on a component.

**Role element – @schemeIdUri attribute**
This attribute specifies the scheme used for the role value. The DASH specification itself defines a simple role scheme (URI: urn:mpeg:dash:role:2011) which shall be used here.

**Role element – @value attribute**
This attribute specifies the actual role value. Regarding audio description, the applicable values depend on the actual purpose of the audio description. The following values from the mentioned scheme are used here:

- alternate (indicating Broadcast mix AD)
- commentary (indicating Receiver mix AD)

#### 3.2.2.2 Accessibility element

**Location:**
AdaptationSet/ContentComponent elements - as part of a MPD

**Defined in:**

- ISO/IEC 23009-1:2014 [4], § 5.8.4.3
- ETSI TS 103 285 V1.1.1 [5], § 6.1.2
- ETSI TS 102 822-3-1 V1.9.2 [6], § A.15

**Description:**
The Accessibility element is a generic element in DASH to specify an arbitrary accessibility property that is based on a specific scheme. The element can be used multiple times on a component.

**Accessibility element – @schemeIdUri attribute**
This attribute specifies the scheme used for the accessibility value. DASH itself does not define a scheme here. For audio description, the *Audio Purpose* scheme from the *TV Anytime* project is used (URI:urn:tva:metadata:cs:AudioPurposeCS:2007).

**Accessibility element – @value attribute**
This attribute specifies the actual accessibility value.

To indicate audio description, the value 1 from the mentioned scheme is used here.

#### 3.2.2.3 SubRepresentation element

**@mimeType attribute**

**Location:**
AdaptationSet/Representation/SubRepresentation elements - as part of an MPD.

**Defined in:**

- ISO/IEC 23009-1:2014 [4], § 5.3.7.2
- ETSI TS 103 285 V1.1.1 [5], § 6.1.1

**Description:**
The mandatory mimeType attribute is used to describe the MIME type of the actual media segments.

For audio description, the value "audio/mp4" is used for all audio codecs mentioned in the audio chapter (ETSI TS 103 285 V1.1.1, §6).

**@codecs attribute**

**Location:**
AdaptationSet/Representation/SubRepresentation elements - as part of an MPD.

**Defined in:**

- ISO/IEC 23009-1:2014 [4], § 5.3.7.2
- ETSI TS 103 285 V1.1.1 [5], § 6.1.1

**Description:**
The mandatory codecs attribute specifies the used codec. It shall also include profile and level information, if applicable. For audio description, the value depends on the audio codec actually used and starts with a matching codec identifier.

#### 3.2.2.4 ProgramInformation element

**@lang attribute**

**Location:**
AdaptationSet/ContentComponent/ProgramInformation elements - as part of an MPD.

**Defined in:**

- ISO/IEC 23009-1:2014 [4], § 5.3.3.2
- ETSI TS 103 285 V1.1.1 [5], § 6.1.2

**Description:**
The lang attribute is used to describe the actual language of e.g. an audio description adaptation set. Its value is specified according to RFC 5646 [18].

#### 3.2.2.5 Representation element

**@dependencyId attribute**

**Location:**
Representation element - as part of a MPD.

**Defined in:**

- ISO/IEC 23009-1:2014 [4], § 5.3.5.2
- ETSI TS 103 285 V1.1.1 [5], § 6.1.2

**Description:**
The dependencyId attribute, if present, indicates that a representation is a dependent representation, and hence depends on other representations in order to be rendered. The attribute value consists of a space-separated list of one or more representations (actually, their IDs) this representation depends on.

Regarding audio description this element is only used for receiver-mix AD.

### 3.2.3 Proposed solutions for ImAc services

The ImAc project identified the following features that are not sufficiently supported by relevant standards:

- Signalling of Ambisonics
- Description of audio content variations

The following subsections describe possible solutions to realize these features. The proposals contain both standard extensions and the specific solutions used in ImAc.

#### 3.2.3.1 Signalling of Ambisonics in MPEG DASH

For ImAc's spatial audio services, audio will be delivered as Ambisonic audio (1st order) using generic audio compression formats, in our case AAC. This may be a common use case, since Ambisonic decoder libraries are widely available, and there are only few alternatives currently available to realize spatial audio on a wide range of client devices. We realized that MPEG-DASH is currently missing the possibility of signalling this at the MPD level, such that a player would be allowed to pick the right set of audio streams according to its capabilities.

Another use case is the possibility to have at the same time an Ambisonic representation and a mono/stereo non spatialized track (usually called head-locked). This sound data could be carried in a dedicated representation or embedded within the stream containing the Ambisonic data. For the latter case, an extension of the channel mapping would be needed to signal either a mono channel or a stereo pair (left / right).

**Proposal**

We proposed at MPEG (m43552) the addition of a new descriptor for Ambisonic, covering our use case. In particular, our proposal was to define an AmbisonicChannelMap descriptor with the following properties:

*urn:* urn:mpeg:dash:ambi-map:2018

*value syntax:* a comma separated list of either integer or 'L','R' or 'M'

Integers specify the Ambisonic Channel Number (ACN), 'M' indicates a mono channel, 'L' a left and 'R' a right channel.

*semantics:* The value field gives the mapping between the channels present in the stream and their corresponding ACN or mono/left/right ordered list of channels. 'M','L','R' indicate "head-lock" channels which shall not be spatialized, but still mixed in the output.

*position:* when present, this descriptor shall be either at the AdaptationSet or Representation level in the MPD.

*optionality:* this descriptor may be signaled as a Supplemental property if the stream contains only the first order Ambisonic channel, or the mono head-lock or the full stereo pair head-lock. Otherwise, it shall be signaled as an Essential property.

Annex I provides an example of the proposed implementation for this feature, as an MPD snippet.

#### 3.2.3.2 Audio variations

The ImAc audio services rely on the delivery of several audio streams for each service. A service in this context is for example Audio Description mixed with the main audio in one language. Each stream contains a different audio mix and/or different audio treatment. We must signal the different treatments to the player application such that it can pick the intended stream for playback.

ImAc has introduced two parameters for the project's purposes:

- AD Gain – The gain of the AD part of the signal in the mix (when using broadcast-mix mode).
- Audio mode – Spatial audio treatment, e.g. used to spatially position AD segments in the (audio) scene. [2]
- Audio narrative - Described different types of narrative approaches (descriptive metadata).

The use case to identify custom parameters on the level of the streaming format, like MPEG DASH in our case, is not restricted to this project, but a general need when introducing features as the above mentioned.

Please note that the approach described above assumes a server-side audio mixing. Alternative approaches using client-side mixing are possible as well and might be a good option depending on the use case. We will not discuss the advantages and disadvantages of server-side and client-side mixing here. DASH does support the signalization of audio sources that need to be mixed by the client (called Receiver mix in DASH, refer to 2.2.1 for details).

[2] The audio modes are explained in Deliverable D5.2, second iteration in chapter 3.1.1.4 [3]

**Proposal**

Our proposal consists of using a custom namespace attribute to carry the necessary information.

In particular, two custom attributes have been proposed for its addition to the DASH manifest (i.e. the MPD). The separate namespace was defined to avoid conflicts in player applications. The attributes are added to the Representation element under the namespace "http://www.imac-project.eu" with the abbreviation "imac". The two new attributes are:

**@gain**

Values: "low", "medium", "high"

**@mode**

Values: "classic", "static", "dynamic"

A sample implementation of this feature in ImAc is provided in Annex I.

## 3.3 Spoken subtitles

### 3.3.1 Motivation

Spoken subtitles (also called audio subtitles – AST) can be realized in two ways: either by using a text-to-speech engine at the client side, or by delivering the spoken subtitles as audio to the client.

The realization in the ImAc project is done by delivering audio, due to some of the tested features that require audio rendering process that can better be done at broadcaster premises, without facing the challenge for real time rendering. In this scenario, subtitles are delivered similarly to the Audio Description service. The main difference in the ImAc implementation is that the AST is mixed at the client side (receiver-mix), where all AD variations are pre-rendered before delivery (broadcast-mix).

In section 2 we provided a gap analysis regarding the signalling of ImAc services and an overview of the required ImAc extensions. In this section, we describe possible solutions in detail, based on existing standards. Section 4.2 describes relevant data for audio and audio description in particular in the state-of-the-art MPEG DASH specification. The spoken subtitle service is similar, but not identical to audio description. Standardized signalling data for audio description could be used for spoken subtitles as well (please refer to section 4.2 for state-of-the-art signalling for AD).

The requirements that arise from the ImAc services are:

1. Means to signal audio as spoken subtitle track in MPEG DASH.
2. Means to signal custom properties of audio streams in MPEG DASH.

A proposal for a possible signalling of AST is described in section 5.2. The signalling of audio properties (required to realize the ImAc audio modes) is equal to the signalling for audio description and described in section 4.3.2.

### 3.3.2 Proposed solutions for ImAc services

#### 3.3.2.1 Signalling of spoken subtitles

As described in section 5.1, the player must be able to identify AST streams correctly in order to present the user a corresponding choice for the available services, in this case for AST. To ensure a correct rendering, the audio tracks that carry the AST must be identified as AST, and it must be signalled if this audio contains AST only (for receiver-mix) or AST mixed with the main audio (broadcast-mix). The latter can be signalled by DASH as described in section 4.2.1 using the Role element, together with the values “alternate” for broadcast mix AST and “commentary” for receiver mix AST.

Identifying audio as AST is not possible by the state-of-the-art DASH standard.

**Proposal**

Following the above standing requirement, we propose to signal AST audio at the DASH layer (or equivalent when using other streaming formats, like HLS [14] or CMAF [15]).

For the usage of MPEG DASH, the Accessibility element should be used to carry this information. This element is intended to signal the type of access service which is our intention here.

We suggest adding an audio-subtitles identifier as a new value to the descriptor “urn:tva:metadata:cs:AudioPurposeCS:2007” for the Accessibility element. This descriptor uses integer values that are mapped to target audience groups. The value “1” is used to indicate the target group “for the visually impaired”, which is also used to signal audio description. The “visually impaired” users are also the target group of the AST service, and an Accessibility element should be set accordingly. A second Accessibility element should carry the concrete type of the access service (AD or AST).

## 3.4 Sign Language

### 3.4.1 Motivation

The sign language video (SL) is presented in a similar manner than the subtitles in the ImAc project. That means that the SL video is shown at a certain position on the current viewport and stays at that position fixed-to-screen.

As for ST, we have developed enhanced presentation options for the SL video, which are listed below:

- The SL video is enriched with cues that indicate the speaker location within the scene. This is done using an arrow indicator that is shown under the video, pointing in the direction of the related speaker in the scene, who is currently being interpreted. The presentation of a radar to indicate the position of the target speaker is also supported.
- The speaker’s name, or a textual description, can be provided below the SL video to help the viewer with the speaker identification.
- It is possible to dynamically hide the signer video, when the SL interpreter is currently inactive, i.e. not translating anybody, and to show it when the speaker is again active.

All these options require time-related metadata to be provided in addition to the video.

In addition, the presentation of SL video fixed to the user’s viewport, and to possibility to customize its presentation in terms of size, position and language make necessary the provision of an independent stream as the main 360º video stream, as a burned-in video would not enable these features.

**Identified gaps in state-of-the-art standards**

We found that current standards and specifications do not provide the required possibilities to realize all features of the developed presentation options. In this section, we describe possible solutions in detail, based on existing standards.

In particular, we have identified the following requirements that arise from the ImAc use cases[3]:

1. Means to signalize the sign language video intended size and position on screen
2. Means to signalize the 360° position of the speaker being interpreted
3. Means to signalize speaker identification information, like speaker's name and/or a color code.
4. Means to signalize whether the sign language interpreter is currently active or inactive

Section 6.2 describes relevant data in the state-of-the-art MPEG DASH specification for the correct signalling of the SL video within DASH.

Proposals for possible solutions to meet these requirements, based on DASH and TTML, are described in Section 6.3.

## 3.4.2 State of the art

### 3.4.2.1 Role element

**Location:**

AdaptationSet/ContentComponent elements - as part of an MPD.

**Defined in:**

- ISO/IEC FDIS 23009-1 [16], § 5.8.4.2, § 5.8.5.5
- DASH-IF Interoperability Points V4.2, [17] § 3.9.2

**Description:**

The Role element is an element in DASH to specify an arbitrary role for the affected component that is based on a specific scheme. The element can be used multiple times on a component.

**Role element – @schemeIdUri attribute**

This attribute specifies the scheme used for the role value. The DASH specification itself defines a simple role scheme (URI: urn:mpeg:dash:role:2011) which should be used here.

**Role element – @value attribute**

This attribute specifies the actual role value.

Regarding sign language, the following values from the mentioned scheme are used here:

- sign

### 3.4.2.2 Accessibility element (draft status)

**Location:**

AdaptationSet/ContentComponent elements - as part of an MPD

**Defined in:**

- ISO/IEC FDIS 23009-1 [16], § 5.8.4.3, § 5.8.5.5
- DASH-IF Interoperability Points V4.2 [17], § 3.9.2, § 3.9.4.5

[3] The use cases that were tested with end users during various test phases are summarized in D5.2, second iteration [3]. Features for SL are listed in chapter 1.1.1.3.

**Description:**
The Accessibility element is a generic element in DASH to specify an arbitrary accessibility property that is based on a specific scheme. The element can be used multiple times on a component.

**Accessibility element – @schemeIdUri attribute**
This attribute specifies the scheme used for the accessibility value. DASH itself doesn't define a scheme here. However the mentioned role scheme (URI: urn:mpeg:dash:role:2011) is used here, too.

**Accessibility element – @value attribute**
This attribute specifies the actual accessibility value.

- Regarding sign language, the following values from the mentioned scheme are used here: sign

#### 3.4.2.3 AdaptationSet element of metadata file

**Location**
MPD/Period/AdaptationSet

**Attribute @id**
The @id attribute must be set on this AdaptationSet in order to identify it from the AdaptationSet that contains the signer video. The value can be arbitrary of type xs:string.

**Attribute @contentType**
The @contentType attribute defines the AdaptationSet content as "application". This is used since MPEG DASH specifies to use "application" in case the AdaptationSet content does not relate to any of the other media content types. (refer to RFC4288, $4.2 [13])

**Attribute @mimeType**
The @mimeType specifies the media content as TTML using the value "application/ttml+xml".

**Role element - @schemeIdUri attribute**
A Role element is added to the AdaptationSet, using a user defined descriptor which identifies the AdaptationSet purpose as a signer metadata track. The ImAc descriptor used for other purposes in the project is adopted for this as well: urn:imac:access-identifier:2019

**Role element - @value attribute**
An additional value is defined particularly for the identification of a signer metadata track: sign-metadata.

A sample implementation of the signaling for the SL video in ImAc is provided in Annex I.

### 3.4.3 Proposed solutions for ImAc services

The use case we have in ImAc requires time related metadata for the SL video. This is not supported in current standards.

**Sidecar TTML file to provide metadata for the SL service**
For a flexible and comprehensive carriage of all required time-related metadata for the SL service, the usage of a sidecar metadata file is proposed. In particular, the decision has been to use a TTML file with the same format as for the IMSC extensions for 360º. This metadata file contains all required parts / elements for indicating the features required for the SL service: different styles for different speakers; initial and end times for each text block; possibility to add the speakers' names as text blocks; and adding the angles information just as for subtitles. This also has the advantage of allowing the re-usage of the IMSC subtitles rendering module.

As described in Section 6.1, the signer presentation options require time related metadata along with the signer video. Even if the signer video is transmitted as a continuous stream, it can be divided into logical segments, each segment containing a sign language interpreted snippet. The segmentation is only conceptional and will be realized via the TTML sidecar file; the video stream itself does not contain information regarding this segmentation.

The on/off periods in this TTML metadata file can be used to hide the SL video when the interpreter is not active. The SL video can be just a continuous video, or a concatenation of different video segments produced using an SL editing tool. In this second case, the ImAc player is provided with a continuous video with inserted back frames for the off periods. In any case, the information from the metadata file can be used to enable the dynamic presentation of SL, based on the speakers', and thus interpreter's, activity. A segment is represented by a tt:p element in the TTML file.

The sidecar TTML metadata files are signalled in DASH as a separate AdaptationSet of the MPD. The particular signaling solution, together with the main attributes of this file are provided in Annex I.

However, the attributes of this new AdaptationSet, and the particular sections and fields of the metadata file for SL, are detailed in the next subsections for better clarity.

#### 3.4.3.1 Signer video size and position

At this stage, no suggestions are made by the ImAc project regarding handling the size and position of the signer video within the 360° scene.

**ImAc implementation:**

The SL video size and its horizontal position is set by the user via personalization features of the player UI. The vertical position is fixed by the player implementation. Just in case that subtitles and sign language are enabled at the same time (supported feature by the player), if the subtitles position is set to top, then the SL window will be also moved to top, as subtitling is considered the master service in such a case.

In addition, the ImAc player enables the user to reposition all visual elements on screen as desired, thanks to an implemented drag & drop functionality.

#### 3.4.3.2 Location of the speaker

For each segment of the SL interpreter, the horizontal direction of the related speaker needs to be identified.

**Proposed solution**

We propose to re-use the defined attributes for the ST service for such a purpose (see Section 3). In particular, the `imac:equirectangularLongitude` and `imac:equirectangularLatitude` attributes, as part of the imac namespace, are added to the `</tt:p>` elements.

#### 3.4.3.3 Speaker identification information

For each segment of the SL interpreter, information on the related speaker needs to be described.

**Proposed solution**

The data used for speaker identification in the ImAc implementation contains:

- A color code
- The speaker's name

This solution was proved to be preferred to users than providing emojis / pictograms to represent speakers' faces, in conducted user tests.

As for the previous feature, the IMSC sections and fields are re-used. In particular, the color code and styling information can be provided in the `<tt:styling>` section of the TTML file, and the speakers' names can be provided in `<tt:p>` elements, as if they were subtitle text blocks.

#### 3.4.3.4 Status of sign language interpreter (active / inactive)

The sign language interpreter status describes when a SL interpreter is active. When the interpreter is inactive, the signer video may show him/her in a neutral position, or the video may show any other placeholder between the last and the next sign language segment, or directly the video window can be hidden (which can contribute to a higher immersion). If the produced SLe stream is discontinuous, then a continuous stream can be delivered by inserting black frames for the inactive periods.

**Proposed solution**

As for the previous features, the IMSC sections and fields are re-used. In particular, the status and presence of an active SL segment will be determined / signalized by using `<tt:p>` elements, and the on/off periods will be determined by their `begin` and `end` attributes of each of these elements.

Annex I provides an example implementation of how all the features described in this section are provided in ImAc.

# Annex I. Sample Implementations in ImAc

## Subtitles for 360°

The ImAc implementation for subtitles is based on the subtitle format IMSC and custom attributes that are used to indicate speaker positions in the scene. The attribute imac:audioSourceAzimuth describes the azimuth and the attribute imac:audioSourceElevation describes the elevation of the direction in the scene where the speaker is located.

The center point [0; 0] is defined as the point on the sphere where the center of the equirectangular video texture is projected. Turning left results in increasing (positive) longitude values. Turning right results in decreasing (negative) longitude values.

Below is a sample snippet from an IMSC file:

```
<tt:p
       xml:id="s1"
       region="bottom"
       style="defaultStyle"
       begin="00:00:01.000"
       end="00:00:04.000"
       imac:audioSourceAzimuth="-30"
       imac:audioSourceElevation="20">

       <tt:span
       style="colorYellow">
       Sample subtitle
       </tt:span>

</tt:p>
```

## Easy-to-Read Subtitles for 360º

The ImAc implementation for Easy-to-Read subtitles is based on the definition of a custom descriptor for the accessibility element, such that no conflicts with the existing standards occur. It is intended to keep this custom solution in place until a standardized solution becomes available. Accordingly, the signalling of easy-to-read subtitles in ImAc is as follows:

- Accessibility element with schemeIdUri “urn:imac:access-identifier:2019” and value “easy-to-read”.

Next, an example of an MPD snippet including the signalling of easy-to-read subtitles, as implemented in ImAc, is provided:

```
<AdaptationSet …>

<Role

       schemeIdUri="urn:mpeg:dash:role:2011"

       value="alternate"/>

<Accessibility

       schemeIdUri="urn:imac:access-identifier:2019"
```

```
        value="easy-to-read"/>
<Representation ...>
        <BaseURL></BaseURL>
</Representation>
</AdaptationSet>
```

## Signaling of Ambisonic Audio in MPEG DASH

ImAc has proposed different alternatives to signalized Ambisonics audio in the MPD.

Backward compatible order 0

```
<AS id="1">
    <SupplementalProperty schemeIdUri="urn:mpeg:dash:ambi-map:2018" value="0"/>
    <Representation/> //1 channel audio
</AS>
<AS id="2">
    <EssentialProperty schemeIdUri="urn:mpeg:dash:ambi-map:2018" value="1 2 3"/>
    <Representation/> //3 channels audio
</AS>
<Preselection preselectionComponents='1 2'/>
```

Order 1 with head-lock

```
<AS id="2">
    <EssentialProperty schemeIdUri="urn:mpeg:dash:ambi-map:2018" value="L R 0 1 2
3"/>
    <Representation> //6 channels audio
</AS>
```

Order 1 with backward compatible head-lock

```
<AS id="1" lang="en">
    <SupplementalProperty schemeIdUri="urn:mpeg:dash:ambi-map:2018" value="L R"/>
    <Representation> //2 channels audio
</AS>
<AS id="3" lang="fr">
    <SupplementalProperty schemeIdUri="urn:mpeg:dash:ambi-map:2018" value="L R"/>
    <Representation> //2 channels audio
</AS>
```

```
<AS id="4">
    <EssentialProperty schemeIdUri="urn:mpeg:dash:ambi-map:2018" value="0 1 2 3"/>
    <Representation> //4 channels audio
</AS>
<Preselection preselectionComponents='1 3' lang="en/>
<Preselection preselectionComponents='2 3' lang="fr/>
```

## Personalized Audio Description for 360º services

The ImAc implementation for personalized AD relies on the use of two custom attributes to be added to the MPD. A separate namespace is defined to avoid conflicts in player applications. The attributes are added to the Representation element, under the namespace "http://www.imac-project.eu" with the abbreviation "imac".

Next, an example of an MPD snippet including the signalling of AD tracks, as implemented in ImAc, is provided. Next that a new AdaptationSet needs to be provided for each AD variant (language, mode, gain).

```
<MPD ... xmlns:imac="namespace:for:imac:audiodescription" ...>
// One AdaptationSet for each AD variant
<AdaptationSet id="eng1" lang="eng" segmentAlignment="true" startWithSAP="1">
        <Role schemeIdUri="urn:mpeg:dash:role:2011" value="alternate"/>
        <Representation ...>
        <imac:AudioDescription gain="low" mode="static"/>
        <AudioChannelConfiguration … value="2"/>
        <SegmentTemplate .../>
        </Representation>
</AdaptationSet>
<AdaptationSet id="eng2" lang="eng" segmentAlignment="true" startWithSAP="1">
        <Role schemeIdUri="urn:mpeg:dash:role:2011" value="alternate"/>
        <Representation ...>
        <imac:AudioDescription gain="medium" mode="dynamic"/>
        <AudioChannelConfiguration … value="2"/>
        <SegmentTemplate .../>
        </Representation>
</AdaptationSet>
```

# Personalized Sign Language presentation for 360º services

The ImAc implementation for personalized SL presentation for 360º videos relies on the addition of a new AdaptationSet for the SL video, to be presented in parallel with the 360º video; and on the addition of a new AdaptationSet for the sidecar TTML file, providing the necessary metadata for the SL video: on/off periods, angles where the speakers are located, and speakers' names and/or textual descriptions relies on the use of two custom attributes to be added to the MPD.

Next, an example of an MPD snippet including the signalling for the SL service, is provided. Note that a new pair of AdaptationSet should be provided for each SL language.

```
// AdaptationSet for the SL video

// ISO 639-3 standard is used for the SL Language codes, as supported by DASH

<AdaptationSet id="signerVideo_ger" lang="gsg">

<SupplementalProperty schemeIdUri="urn:imac:signer-metadata-adaptation-set-id:2019"
value="signerMetadata_ger"/>

<Role schemeIdUri="urn:mpeg:dash:role:2011" value="sign"/>

       <Representation id="signer_with_specific_resolution">

       <BaseURL></BaseURL> //URL of MPD

       </Representation>

</AdaptationSet>

// AdaptationSet for the SL metadata file

<AdaptationSet id="signerMetadata_ger" ... mimeType="application/ttml+xml"
lang="ger">

<Role schemeIdUri="urn:imac:access-identifier:2019" value="sign-metadata"/>

       <Representation id="" bandwidth="1000">

       <BaseURL></BaseURL> //URL of the sidecar TTML file

       </Representation>

</AdaptationSet>
```

Next, a snippet of the sidecar TTML file for the SL service metadata is provided:

```
<tt:head>

// Definition of text styling

<tt:styling>

       <tt:style xml:id="C1" imac:type="stCharacter" tts:fontSize="34px"
       tts:color="#FFFF00" tts:backgroundColor="transparent" tts:padding="5"/>

       <tt:style xml:id="A1" tts:textAlign="left" imac:type="stTextAlign"/>

       ...

</tt:styling>

...

</tt:head>
```

```
<tt:body>

        …

          // p elements with temporal segments for each speaker's interpretation with
          a specific angle settings, begin and end time, and text

          // Note that different consecutive p elements may include information about
          the SL segment, but with varying angles if the speaker moves around the 360°
          scene

            <tt:p    xml:id="p1"    region="R1"    style="A1"    begin="00:00:04.920"
            end="00:00:44.240"                      imac:equirectangularLongitude="300"
            imac:equirectangularLatitude="10">

               <tt:span style="C1">Speaker's Name</tt:span>

            </tt:p>

        ...

</tt:body>
```